\documentclass[a4paper,11pt]{article}
\usepackage{pos}

\title{Progress of the FLASHForward X-2 high-beam-quality, high-efficiency plasma-accelerator experiment}

\ShortTitle{Progress of the FLASHForward X-2 experiment}

\author*[a]{C. A. Lindstr{\o}m}
\emailAdd{carl.a.lindstroem@desy.de}
\affiliation[a]{Deutsches Elektronen-Synchrotron DESY, Notkestraße 85, 22607 Hamburg, Germany}

\author[a,b]{J. Beinortaite}
\emailAdd{judita.beinortaite@desy.de}
\affiliation[b]{University College London, London, United Kingdom}

\author[a]{J. Bj{\"o}rklund Svensson}
\emailAdd{jonas.bjorklund-svensson@desy.de}

\author[a,c,d]{L. Boulton}
\emailAdd{lewis.boulton@desy.de}
\affiliation[c]{SUPA, Department of Physics, University of Strathclyde, Glasgow, United Kingdom}
\affiliation[d]{The Cockcroft Institute, Daresbury, United Kingdom}

\author[b]{J. Chappell}
\emailAdd{james.chappell@desy.de}

\author[a]{J. M. Garland}
\emailAdd{matthew.james.garland@desy.de}

\author[a,e]{P. Gonzalez}
\emailAdd{pau.gonzalez@desy.de}
\affiliation[e]{Universit{\"a}t Hamburg, Luruper Chaussee 149, 22761 Hamburg, Germany}

\author[a]{G. Loisch}
\emailAdd{gregor.loisch@desy.de}

\author[a,e]{F. Pe{\~n}a}
\emailAdd{felipe.pena@desy.de}

\author[a]{L. Schaper}
\emailAdd{lucas.schaper@desy.de}

\author[a]{B. Schmidt}
\emailAdd{bernhard.schmidt@desy.de}

\author[a]{S. Schr{\"o}der}
\emailAdd{sarah.schroeder@desy.de}

\author[a]{S. Wesch}
\emailAdd{stephan.wesch@desy.de}

\author[a]{J. Wood}
\emailAdd{jonathan.wood@desy.de}

\author[a]{J. Osterhoff}
\emailAdd{jens.osterhoff@desy.de}

\author[a]{R. D'Arcy}
\emailAdd{richard.darcy@desy.de}

\abstract{FLASHForward is an experimental facility at DESY dedicated to beam-driven plasma-accelerator research. The X-2 experiment aims to demonstrate acceleration with simultaneous beam-quality preservation and high energy efficiency in a compact plasma stage. We report on the completed commissioning, first experimental results, ongoing research topics, as well as plans for future upgrades.}

\FullConference{
  The European Physical Society Conference on High Energy Physics (EPS-HEP2021)\\
  26-30 July 2021\\
  Online conference, jointly organized by Universität Hamburg and the research center DESY
}

\begin{document}
\maketitle

\section{Introduction}
Plasma-wakefield acceleration is a technique for accelerating charged particles to high energies in distances 10--1000 times shorter than that possible with conventional RF accelerators \cite{Joshi2003,Leemans2009}. Large, multi-GV/m accelerating fields can be sustained in a plasma within a small region trailing either an intense laser pulse \cite{Tajima1979} or particle beam \cite{Veksler1956,Chen1985,Ruth1985}. Beam-driven plasma acceleration is particularly interesting for applications that require high energy and beam power, since beam drivers can drive wakes stably across long plasma-accelerator stages and be produced with high wall-plug efficiency. 

The FLASHForward facility \cite{Aschikhin2016} at DESY is dedicated to research and development of beam-driven plasma accelerators, with a view towards both photon-science applications like free-electron lasers (FELs) and high-energy-physics applications such as a linear collider. Stable, high-current (kA), high-charge (nC), low-emittance (mm-mrad) beams are accelerated to approximately 1 GeV in the FLASH linac \cite{Faatz2016} and delivered to the FLASHForward beamline at up to 10~Hz. Since FLASH operates superconducting RF cavities, bunch trains with {\textmu}s bunch spacing are also available, which opens the unique possibility of plasma acceleration at MHz repetition rate.

Experimentation is mainly divided into three flagship experiments \cite{Darcy2019a}. This includes an internal-injection experiment (X-1), producing high-quality electron bunches directly from the plasma \cite{Deng2019,Knetsch2021}; an external-injection experiment (X-2), accelerating existing bunches from FLASH with high efficiency while preserving their beam quality; and a high-repetition-rate experiment (X-3), investigating beam-driven plasma acceleration at very high repetition rates. This paper reports on the current progress of the X-2 experiment.

\section{Experimental goals}
The overarching goal of the X-2 experiment is to demonstrate operation of a self-consistent plasma-accelerator stage, to be used by itself or as a building block for a multistage accelerator \cite{Steinke2016,Lindstrom2016,Lindstrom2021a}. This requires accurate measurement and tight control of every aspect of the driving and trailing electron bunches, as well as the plasma density profile, in which the acceleration occurs. 

Pioneering experiments at SLAC's FFTB and FACET facilities have already demonstrated high gradients and large energy gain \cite{Blumenfeld2007}, as well as high instantaneous energy-transfer efficiency between the driver and the trailing bunch \cite{Litos2014}. However, several challenges remain to be tackled before beam-driven plasma-accelerator stages are ready for applications. X-2 therefore has the following specific goals: beam-quality preservation, including energy spread and emittance, as well as high \textit{overall} energy efficiency (i.e., transferring a large fraction of the initial driver energy to the trailing bunch), all at an accelerating gradient of at least 1~GV/m.

\subsection{Energy-spread preservation}
Many applications require the beam's energy spread to be 1\% or less; some (FELs) even as low as 0.1\%. This is a challenge in a plasma accelerator due to the short length of the plasma-accelerator cavity, when compared to the length of the bunch, which can span a large range of accelerating phases. Moreover, to extract energy and maintain a high efficiency, the wakefield must be loaded by the beam. In this case, energy-spread preservation is only possible through precise shaping of the trailing-bunch current profile, as described in Ref.~\cite{Tzoufras2008}.

\subsection{Emittance preservation}
The transverse beam quality, specifically the normalized emittance, also needs to be preserved throughout a plasma-accelerator stage, ideally at the mm-mrad level or below. This is possible in the exposed ion channel of a blown out plasma wake as the focusing fields are linear. Preserving emittance also requires the trailing bunch to be transversely aligned with the driving bunch, as well as to have a small (mm-scale) Twiss beta function matched to the strong focusing inside the wake, ideally in both planes. Moreover, measuring the emittance of bunches exiting a plasma is nontrivial due to the very small spot sizes---high resolution imaging of the beam size is required.

\subsection{High overall energy efficiency}
Lastly, high energy-transfer efficiency between the driver and the trailing bunch is not sufficient: the driver simultaneously needs to deplete a significant portion of its initial energy to ensure high overall efficiency. Such driver depletion can be challenging, as it requires stable deceleration of the driver and acceleration of the trailing bunch over long distances. The goal of X-2 is to demonstrate simultaneous 50\% depletion of the total driver energy as well as 50\% of that energy being transferred to the trailing bunch---an overall energy efficiency of 25\%.

\section{Commissioning}
First installations of the FLASHForward beamline started in 2015 and continued until 2018, as described in Ref.~\cite{Libov2018}. Sections for beam extraction, compression and dispersion cancellation, matching and final focusing, as well as post-plasma spectrum diagnostics were completed, which allowed first experiments \cite{Darcy2019b} in a short (33-mm-long) plasma cell. Since 2018, commissioning of hardware has expanded to now include energy collimators, long plasma cells, both a broadband and a high-resolution dipole spectrometer, as well as a state-of-the-art transverse deflection structure for characterizing the temporal structure of the bunches---all of which are essential to the operation of the X-2 experiment.

\subsection{Energy collimators for two-bunch generation}
In order to produce a drive and trailing bunch pair, the beam is chirped in longitudinal phase space and energetically dispersed in the horizontal plane onto a set of three collimators. These include a wedge-shaped notch collimator for splitting the bunches with an adjustable separation, and two block collimators that can limit the length of each bunch. This setup was installed and tested successfully, and is described in full detail in Ref.~\cite{Schroeder2020a}.

\subsection{Deployment of long discharge plasma cells}
\begin{figure}[h]
	\centering\includegraphics[width=0.8\textwidth]{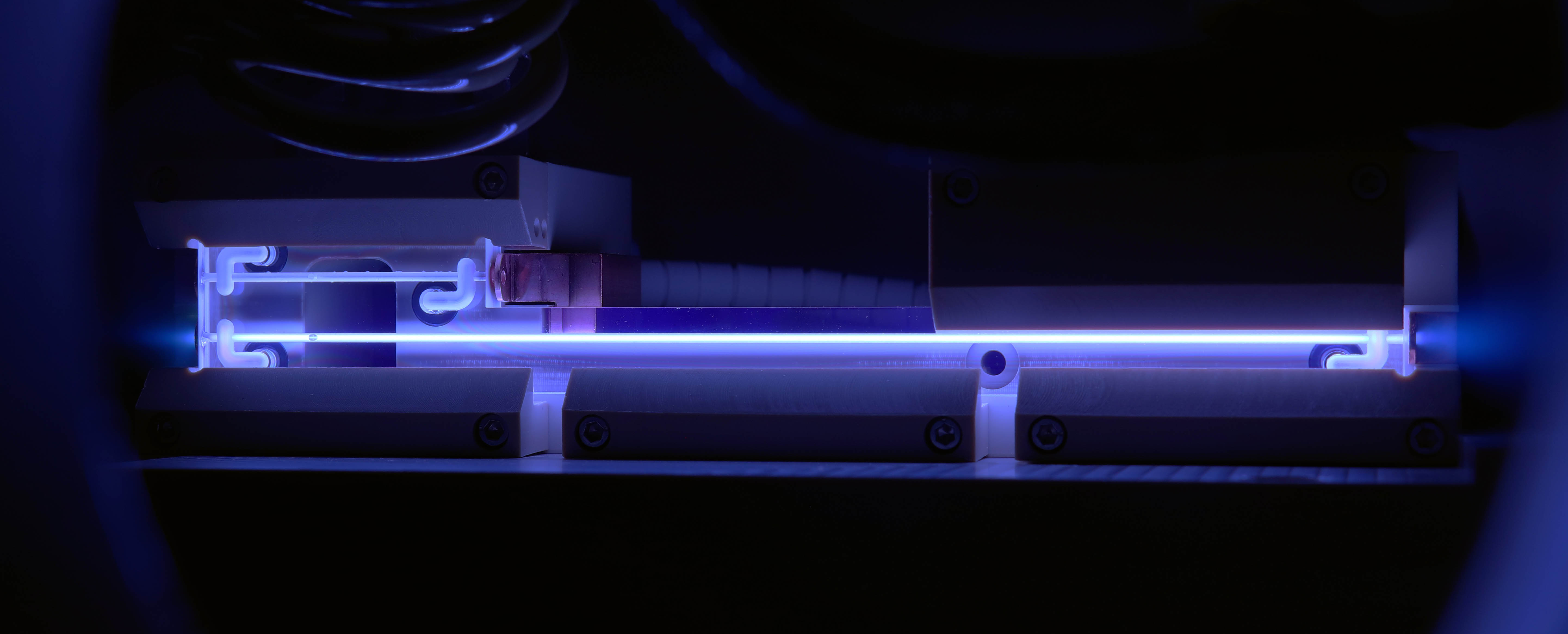}
	\caption{High-dynamic-range image of the dual plasma-cell setup, in which the 195-mm-long capillary is filled with argon gas and discharged with a high-voltage current pulse. Situated above it is a 50-mm-long capillary for setup and precision experiments. In this image, the beam travels from left to right.}
    \label{fig:Fig1}
\end{figure}

A sapphire plasma cell with two capillaries is used for the X-2 experiments, as shown in Fig.~\ref{fig:Fig1}. One capillary is 50 mm long and is used for precision measurements as well as setup and pre-optimization for the longer 195 mm cell, which is used for experiments that require large energy gain and loss. To avoid discharging through the gas inlets, 1.5-m-long spiral-shaped pipes are used (Fig.~\ref{fig:Fig1}; upper left), made out of PEEK plastic to match the ultra-high vacuum (UHV) requirements of FLASH. Argon gas is used for most experiments, as it produces the most stable discharges, and has been operated for more than $10^5$ shots without degrading the capillaries. The cells have also been operated with hydrogen, helium, nitrogen, neon, and krypton. The plasma density is characterized in-situ by observing the broadening of the H-alpha line with an optical spectrometer \cite{Garland2021}. To observe the H-alpha line in argon, 3--5\% hydrogen doping is used.

\subsection{Broadband dipole spectrometer for energy-spectrum measurements}
To measure a depleted driver with up to 100\% energy spread, a broadband, vertically dispersive dipole spectrometer is placed just downstream of the interaction chamber. Two scintillating screens (Kodak LANEX Fine) are placed on the outside surface of the 1-mm-thick dipole vacuum-chamber walls and imaged with six separate cameras. To ensure accurate measurements of the full energy spectrum, detailed energy and charge-density calibrations have been performed, taking into account geometry and lens effects, as well as the temporal response of the scintillating material. With sufficiently intense bunches, a strong saturation effect is observed; this sets an upper limit to the beam intensity of around 5 nC/mm$^2$, requiring non-plasma-interacted beams to be defocused in at least one plane.

\subsection{High-resolution dipole spectrometer for emittance measurements}
Measuring mm-mrad-level emittances of beams emerging from {\textmu}m spot sizes requires large magnetic demagnification and very-high-resolution screens. The broadband spectrometer is limited to a demagnification of $-6$ and a resolution of about 50~{\textmu}m (limited by scattering in the chamber wall). Therefore, another dipole spectrometer was installed just downstream of the broadband spectrometer, which allows larger magnifications (i.e., $-12$) due to a longer drift, as well as an in-vacuum gadolinium aluminium gallium garnet (GAGG) screen imaged by a camera with very-high resolution (using Scheimpflug optics) \cite{Kube2015}. The resolution was measured to be 7~{\textmu}m rms, implying an emittance resolution as low as 0.1~mm~mrad (assuming 8~mm beta functions at the plasma exit).

\subsection{Transverse deflection structure}
A new type of X-band transverse deflection structure (TDS), PolariX \cite{Marchetti2020}, was installed during summer 2019, and is used for resolving the time structure of the short electron bunches. This experimental device is the world's first polarizable TDS, meaning it can be used to streak in any plane, enabling measurement of both horizontal and vertical slice emittances, and critical for producing straight beams (i.e., with no tilt or curvature). The TDS is placed 33~m downstream of the plasma in order to accommodate any future undulators. The time resolution has been measured to be as low as 5~fs, expected to reduce further to 1~fs with future power upgrades. Following the TDS is both a screen in a straight section, for observing streaking in all planes, as well as another screen following a horizontally dispersive dipole magnet to allow measurement of the longitudinal phase space.

\section{Experimental results}
While still early in the full arc of X-2 experimentation, several milestones have already been reached as well as published. Due to the difficulty and level of precision required to demonstrate the ultimate plasma stage, experiments have been performed in order of increasing complexity as well as usefulness to later experiments.

\subsection{Fast beam-focus diagnostics using BPMs}
The exact location and size of the focal waist is critical in the setup and operation of a beam-driven plasma accelerator. However, measuring {\textmu}m-scale bunches with in-focus OTR screens or downstream quadrupole scans can be challenging and time consuming. By observing the distribution of the orbit jitter of the beam, measured with high-resolution cavity beam-position monitors (BPMs) upstream and downstream of the interaction chamber, both the location and beta function of the beam waist can be estimated with good accuracy. This novel technique, reported in Ref.~\cite{Lindstrom2020}, provides rapid (few-shot) feedback, enabling online adjustments of the beam-waist position and beta function with mm-scale accuracy---a crucial building block toward beam matching and emittance preservation.

\subsection{Precision wakefield mapping}
The first plasma-based experimental result from X-2 was a high-resolution, time-resolved measurement of a beam-driven wakefield. Exploiting the near-linear chirp in longitudinal phase space, the high-energy (tail) collimator could be used to sample the wake at 15~fs resolution at percent-level accuracy, as reported in Ref.~\cite{Schroeder2020b}. Comparing the measurement with particle-in-cell simulations required a detailed 6D reconstruction of the bunch, combining the BPM-based technique with TDS measurements and quadrupole scans, as well as longitudinally resolved plasma-density measurements---a process that resulted in a greatly increased understanding of many subtleties of the FLASHForward setup. The technique is now routinely used to optimize the wakefield.

\subsection{Energy-spread preservation with optimal beam loading}
A major milestone for X-2 was achieved with an experiment that demonstrated wakefield flattening via optimal beam loading \cite{Tzoufras2008}, resulting in simultaneous preservation of per-mille-level energy spreads and high instantaneous energy-transfer efficiency ($42\pm4$\%). Charge preservation (100~pC) was also demonstrated. This result was reported in Ref.~\cite{Lindstrom2021b}. The optimized working point was found using a multi-hour (12675-shot) 3D parameter scan, only made possible by the combination of both highly stable electron bunches from FLASH and a highly stable plasma source.

\subsection{Energy doubling}
\begin{figure}[h]
	\centering\includegraphics[width=0.8\textwidth]{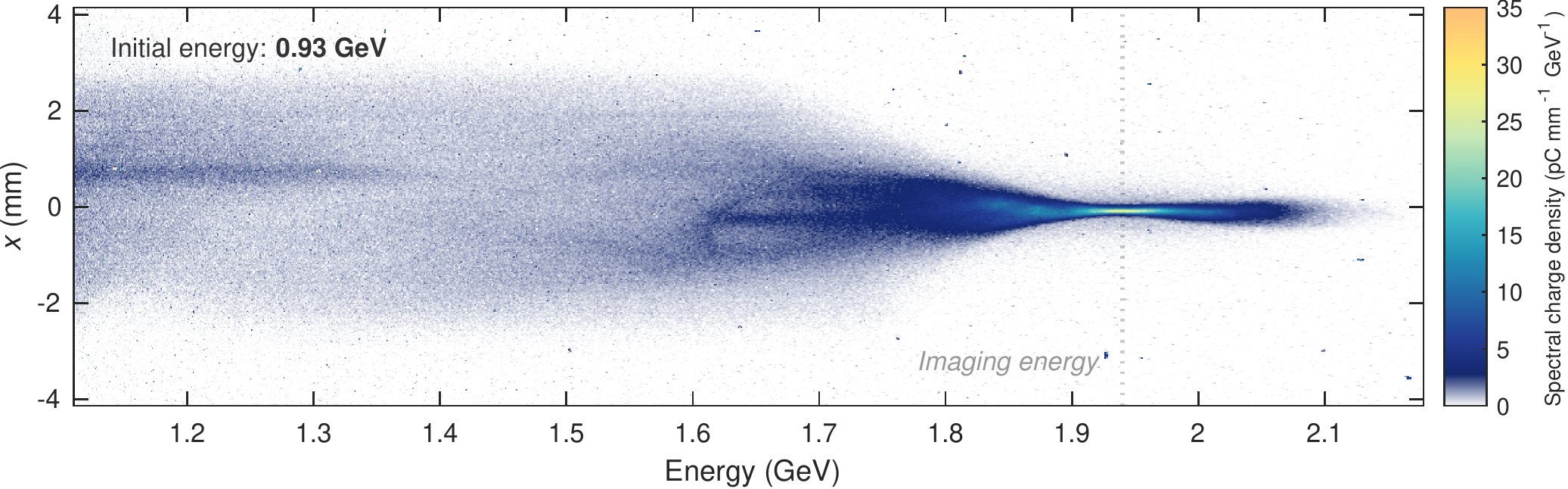}
	\caption{Calibrated spectrometer image of energy-doubled particles starting from an initial energy of 0.93~GeV. The beam was point-to-point imaged from the plasma exit to the LANEX screen at an energy of 1.94 GeV (right side), which leads to strong defocusing and loss of particles at lower energy (left side).}
    \label{fig:Fig2}
\end{figure}

Another milestone was the energy doubling of particles from 1 to 2 GeV in less than 20 cm, as shown in Fig.~\ref{fig:Fig2}. This experiment required stable acceleration over the full length of our longest cell (see Fig.~\ref{fig:Fig1}), and a sustained gradient of at least 5~GV/m. While this is little more than a reproduction of SLAC's landmark result from 2007 \cite{Blumenfeld2007}, it forms the basis of future experiments reaching high overall energy efficiency.

\section{Outlook}
Looking beyond the results described above, a number of experiments are currently in progress. This includes work towards (i) demonstration of emittance preservation, (ii) demonstration of slice-energy-spread preservation, (iii) advanced acceleration-efficiency diagnostics, and (iv) driver-depletion experiments investigating the dynamics of beam particles that reach zero energy. Further experiments are planned for the near future, including investigations of ion motion \cite{Rosenzweig2005} and seeding of the hosing instability \cite{Whittum1991,Mehrling2017}.

2022 will be a year of shutdown for FLASHForward, with planned installation of new and upgraded hardware. Most relevant to X-2 is a new and larger interaction chamber that will house longer plasma cells (e.g., 500 mm long \cite{Turner2021}), in order to facilitate more controlled depletion and acceleration. The new chamber will also host a plasma lens downstream of the accelerator cell, able to operate in both the active \cite{vanTilborg2015,Lindstrom2018} and passive \cite{Chen1989,Ng2001} plasma-lensing regimes. The aim is to tackle the problem of chromatic beam capture, such that the accelerated bunch can be transported downstream to potential future undulators without significant loss of beam quality.

\section{Conclusions}
After a multi-year effort by a large team of people, the X-2 experiment is entering a state of science production. Several milestones have already been achieved, and further results and experiments are in preparation. While much work remains before the ultimate plasma-accelerator stage can be demonstrated, we believe the progress so far provides a solid foundation for the future.

\acknowledgments
The authors would like to thank M. Dinter, S. Karstensen, S. Kottler, K. Ludwig, F. Marutzky, A. Rahali, V. Rybnikov, A. Schleiermacher, the FLASH management, and the DESY FH and M divisions for their scientific, engineering and technical support. This work was supported by Helmholtz ARD and the Helmholtz IuVF ZT-0009 programme.

\end{document}